\documentclass[sn-mathphys-num]{sn-jnl}

\usepackage{graphicx}%
\usepackage{multirow}%
\usepackage{amsmath,amssymb,amsfonts}%
\usepackage{amsthm}%
\usepackage{mathrsfs}%
\usepackage[title]{appendix}%
\usepackage{xcolor}%
\usepackage{textcomp}%
\usepackage{manyfoot}%
\usepackage{booktabs}%
\usepackage{algorithm}%
\usepackage{algorithmicx}%
\usepackage{algpseudocode}%
\usepackage{listings}%
\usepackage{subfigure}
\usepackage{siunitx}
\usepackage{ragged2e} 
\usepackage{booktabs,makecell, multirow, tabularx}
\usepackage{graphicx} 
\usepackage{soul, color, xcolor}
\soulregister\cite7
\soulregister\ref7
\soulregister\textit7
\soulregister\ang7

\begin{document}

\title[Article Title]{Ground calibration and network of the first CATCH pathfinder}

\author[1,2]{\fnm{Yiming} \sur{Huang}}
\author[1,2]{\fnm{Jingyu} \sur{Xiao}}
\author[1]{\fnm{Lian} \sur{Tao}}\email{taolian@ihep.ac.cn}
\author[1,2]{\fnm{Shuang-Nan} \sur{Zhang}}\email{zhangsn@ihep.ac.cn}
\author[1]{\fnm{Qian-Qing} \sur{Yin}}\email{yinqq@ihep.ac.cn}
\author[1]{\fnm{Yusa} \sur{Wang}}
\author[1]{\fnm{Zijian} \sur{Zhao}}
\author[3]{\fnm{Chen} \sur{Zhang}}
\author[1,2]{\fnm{Qingchang} \sur{Zhao}}
\author[1]{\fnm{Xiang} \sur{Ma}}
\author[1,2]{\fnm{Shujie} \sur{Zhao}}
\author[4]{\fnm{Heng} \sur{Zhou}}
\author[1]{\fnm{Xiangyang} \sur{Wen}}
\author[1]{\fnm{Zhengwei} \sur{Li}}
\author[1]{\fnm{Shaolin} \sur{Xiong}}
\author[1]{\fnm{Juan} \sur{Zhang}}
\author[5]{\fnm{Qingcui} \sur{Bu}}
\author[6]{\fnm{Jirong} \sur{Cang}}
\author[7]{\fnm{Dezhi} \sur{Cao}}
\author[4]{\fnm{Wen} \sur{Chen}}
\author[4]{\fnm{Siran} \sur{Ding}}
\author[3]{\fnm{Yanfeng} \sur{Dai}}
\author[1]{\fnm{Min} \sur{Gao}}
\author[4]{\fnm{Yang} \sur{Gao}}
\author[1]{\fnm{Huilin} \sur{He}}
\author[8]{\fnm{Shujin} \sur{Hou}}
\author[9]{\fnm{Dongjie} \sur{Hou}}
\author[10]{\fnm{Tai} \sur{Hu}}
\author[1,2]{\fnm{Guoli} \sur{Huang}}
\author[1]{\fnm{Yue} \sur{Huang}}
\author[6]{\fnm{Liping} \sur{Jia}}
\author[11]{\fnm{Ge} \sur{Jin}}
\author[10]{\fnm{Dalin} \sur{Li}}
\author[4]{\fnm{Jinsong} \sur{Li}}
\author[1,2]{\fnm{Panping} \sur{Li}}
\author[1,2]{\fnm{Yajun} \sur{Li}}
\author[1]{\fnm{Xiaojing} \sur{Liu}}
\author[1,2]{\fnm{Ruican} \sur{Ma}}
\author[1]{\fnm{Lingling} \sur{Men}}
\author[6]{\fnm{Xingyu} \sur{Pan}}
\author[1]{\fnm{Liqiang} \sur{Qi}}
\author[1]{\fnm{Liming} \sur{Song}}
\author[4]{\fnm{Xianfei} \sur{Sun}}
\author[12]{\fnm{Qingwen} \sur{Tang}}
\author[4]{\fnm{Liyuan} \sur{Xiong}}
\author[4]{\fnm{Yibo} \sur{Xu}}
\author[1]{\fnm{Sheng} \sur{Yang}}
\author[1]{\fnm{Yanji} \sur{Yang}}
\author[4]{\fnm{Yong} \sur{Yang}}
\author[1]{\fnm{Aimei} \sur{Zhang}}
\author[10]{\fnm{Wei} \sur{Zhang}}
\author[1,2]{\fnm{Yifan} \sur{Zhang}}
\author[4]{\fnm{Yueting} \sur{Zhang}}
\author[3]{\fnm{Donghua} \sur{Zhao}}
\author[1,2]{\fnm{Kang} \sur{Zhao}}
\author[1]{\fnm{Yuxuan} \sur{Zhu}}

\affil[1]{\orgdiv{Key Laboratory of Particle Astrophysics, Institute of High Energy Physics}, \orgname{Chinese Academy of Sciences}, \orgaddress{\city{Beijing}, \postcode{100149}, \country{China}}}

\affil[2]{\orgdiv{University of Chinese Academy of Sciences}, \orgname{Chinese Academy of Sciences}, \orgaddress{\city{Beijing}, \postcode{100149}, \country{China}}}

\affil[3]{\orgdiv{National Astronomical Observatories}, \orgname{Chinese Academy of Sciences}, \orgaddress{\city{Beijing}, \postcode{100101}, \country{China}}}

\affil[4]{\orgname{Innovation Academy for Microsatellites of Chinese Academy of Sciences}, \orgaddress{\city{Shanghai}, \postcode{200135}, \country{China}}}

\affil[5]{\orgdiv{Institut f\"ur Astronomie und Astrophysik, Kepler Center for Astro and Particle Physics}, \orgname{Eberhard Karls Universi\"at}, \orgaddress{\street{Sand 1},\city{\"ubingen}, \postcode{72076}, \country{Germany}}}

\affil[6]{\orgname{Star Detect CO.}, \orgname{LTD.}, \orgaddress{\city{Beijing}, \postcode{100190}, \country{China}}}

\affil[7]{\orgname{Tsinghua University}, \orgaddress{\city{Beijing}, \postcode{100084}, \country{China}}}

\affil[8]{\orgname{Nanyang Normal University}, \orgaddress{\city{Nanyang}, \postcode{473061}, \country{China}}}

\affil[9]{\orgname{National Institute of Metrology}, \orgaddress{\city{Beijing}, \postcode{100029}, \country{China}}}

\affil[10]{\orgdiv{National Space Science Center}, \orgname{Chinese Academy of Sciences}, \orgaddress{\city{Beijing}, \postcode{100190}, \country{China}}}

\affil[11]{\orgdiv{North Night Vision Technology CO.}, \orgname{LTD.}, \orgaddress{\city{Nanjing}, \postcode{211106}, \country{China}}}

\affil[12]{\orgname{Nanchang University}, \orgaddress{\city{Nanchang}, \postcode{330031}, \country{China}}}

\abstract{The Chasing All Transients Constellation Hunters (CATCH) space mission is focused on exploring the dynamic universe via X-ray follow-up observations of various transients. The first pathfinder of the CATCH mission, CATCH-1, was launched on June 22, 2024, alongside the Space-based multiband astronomical Variable Objects Monitor (SVOM) mission. CATCH-1 is equipped with narrow-field optimized Micro Pore Optics (MPOs) featuring a large effective area and incorporates four Silicon Drift Detectors (SDDs) in its focal plane. This paper presents the system calibration results conducted before the satellite integration. Utilizing the data on the performance of the mirror and detectors obtained through the system calibration, combined with simulated data, the ground calibration database can be established. Measuring the relative positions of the mirror and detector system, which were adjusted during system calibration, allows for accurate installation of the entire satellite. Furthermore, the paper outlines the operational workflow of the ground network post-satellite launch. }

\keywords{X-ray astronomy, Ground calibration, Micro Pore Optics, CATCH}

\maketitle

\section{Introduction}\label{sec1}

The Chasing All Transients Constellation Hunters (CATCH) is a space mission aimed at studying the dynamic universe via X-ray follow-up observations of various multi-wavelength and multi-messenger transients~\cite{licatch, xiao2023orbit, huangsimulation}. CATCH intends to deploy an intelligent-controlled constellation comprising multiple X-ray micro-satellites to address the lack of adequate follow-up observation capabilities in the time-domain astronomy era. If necessary, this mission can also be expanded to a multiwavelength constellation, including UV, optical, and infrared satellites. In the baseline plan, these X-ray satellites are categorized into three types, each serving a different scientific purpose. Type-A satellites are used for immediate timing monitoring after target discovery, type-B satellites are deployed for more in-depth timing, imaging, and spectroscopic follow-up observations, and type-C satellites are specifically designed for polarization measurements. Currently, CATCH is in the process of developing a series of pathfinders to verify key technologies. This paper provides a technical overview of the work conducted on the first pathfinder of the type-A satellite (CATCH-1) before launch, including system calibration, establishment of the calibration database, ground network, and satellite integration. 

CATCH-1, alongside the SVOM mission~\cite{gotz2009svom, atteia2022svom}, was launched on June 22, 2024, and entered a near-equatorial circular orbit at about 625\,km altitude. The optical system of CATCH-1 is a lobster eye X-ray telescope consisting of 4$\times$4 Micro Pore Optics (MPOs)~\cite{angel1979lobster, Hudec2022mpo}. Photons passing through the MPO mirror are focused onto the focal plane by grazing incidence reflections at the sidewalls of the micro pores. The point spread function (PSF) of the MPO mirror is cross-shaped, comprising a central focal spot and two mutually perpendicular cross-arms. The MPO mirror on CATCH-1 is designed and manufactured by the National Astronomical Observatories (NAOC) and the North Night Vision Technology Company Limited (NNVT). To prevent saturation from intense sources, CATCH-1 utilizes a 4-pixel Silicon Drift Detector (SDD) array with fast-time readout capability. The primary detector is positioned on the focal spot with on-axis incidence. On each of the two cross-arms, one detector is placed for auxiliary positioning. Additionally, a detector is positioned on the unfocused diffuse spot for background measurement~\cite{huangsimulation}. The SDD detectors are supplied by the manufacturer KETEK, with the model VITUS H50 for the primary detector and VITUS H20 for the remaining three detectors~\cite{KETEK}. The detection energy range of the SDD detector is 0.5--8\,keV, and the time resolution is 4\,$\mu$s~\cite{lidetector}. The target energy range for the scientific detection by CATCH-1 is 0.5--4\,keV. The main goal of CATCH-1 is to conduct in-orbit verification of the detector system, data analysis software, intelligent control system, fast-pointing capabilities, and other space technologies.

The organization of this paper is outlined as follows: Section~\ref{sec2} presents the results of the system calibration, emphasizing the assessment of the effective area and relative positions. Section~\ref{sec3} describes the establishment of the calibration database by using the results from the calibration and the simulation. Section~\ref{sec4} details the ground network for observatory commanding, data retrieval, and processing. Section~\ref{sec5} introduces the full integration of CATCH-1 before launch. Finally, Section~\ref{sec6} provides a brief summary.

\section{System calibration}\label{sec2}

System calibration refers to the joint ground calibration of the two payloads onboard CATCH-1, and each was fixed on a separate movable stage in the vacuum chamber, using X-rays to obtain a performance benchmark for the combination of the mirror system and the detector system~\cite{freyberg2006mpe, burwitz2023svom, GECAM}. This section describes in detail the process of the system calibration before the integration, including the setup used, the process of aligning the MPO mirror and SDD array to determine the optimal relative position, and the approach and results for calibrating the effective area of the MPO mirror recorded by CATCH-1's 4-pixel SDD array.

\subsection{Setup}\label{sub1}

\begin{figure}[ht]%
        \centering 
        \includegraphics[width=0.65\textwidth]{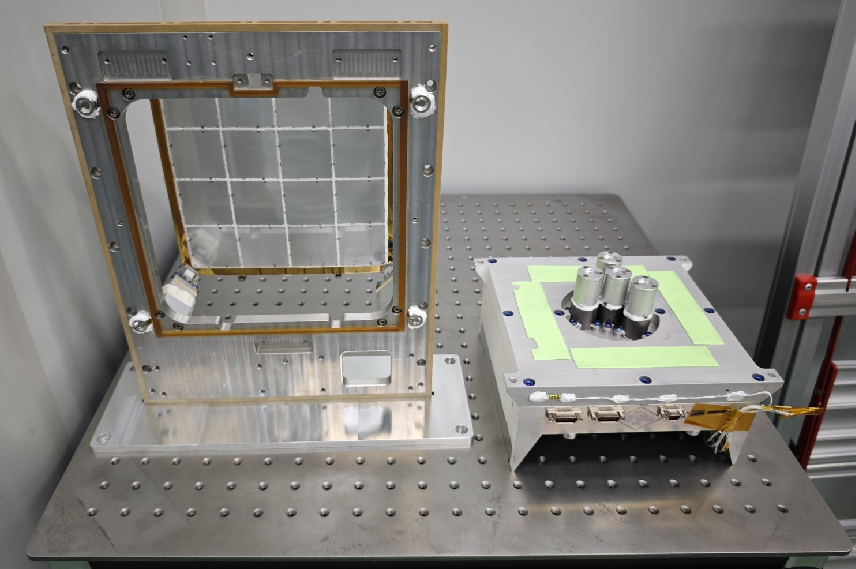}
        \caption{Flight modules of the MPO mirror and the SDD array on CATCH-1.}
        \label{FigSystem}
\end{figure}

\begin{figure}[ht]%
\flushleft
    \begin{minipage}{.44\columnwidth}
        \centering
        \subfigure{\includegraphics[width=0.85\columnwidth]{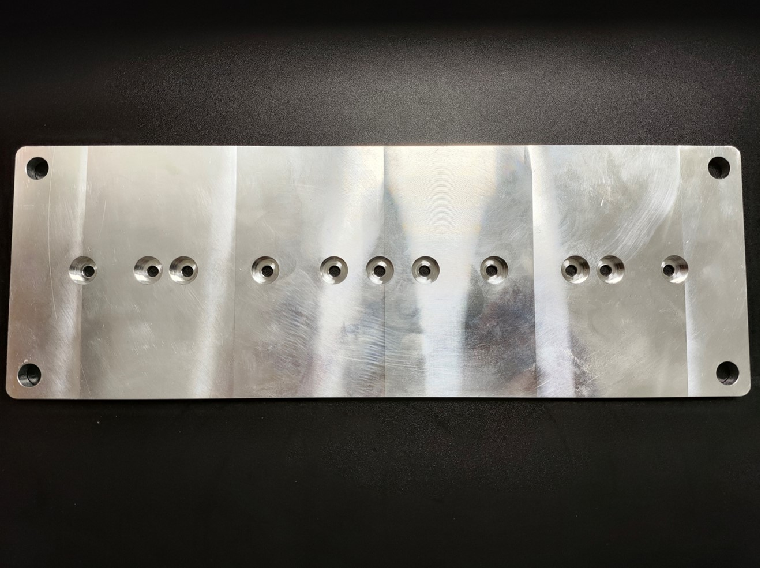}}\\
        \vspace{0.02cm}
        \centering
        \subfigure{\includegraphics[width=0.85\columnwidth]{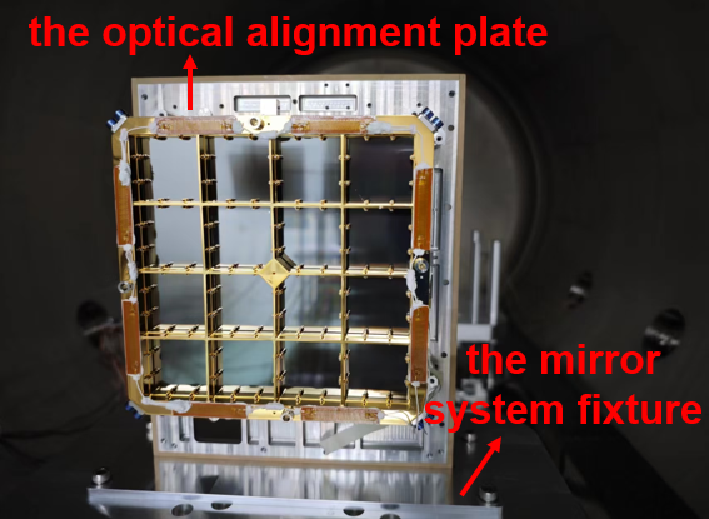}}
    \end{minipage}   
    \begin{minipage}{.45\columnwidth}
        \centering
        \subfigure{\includegraphics[width=1.05\columnwidth]{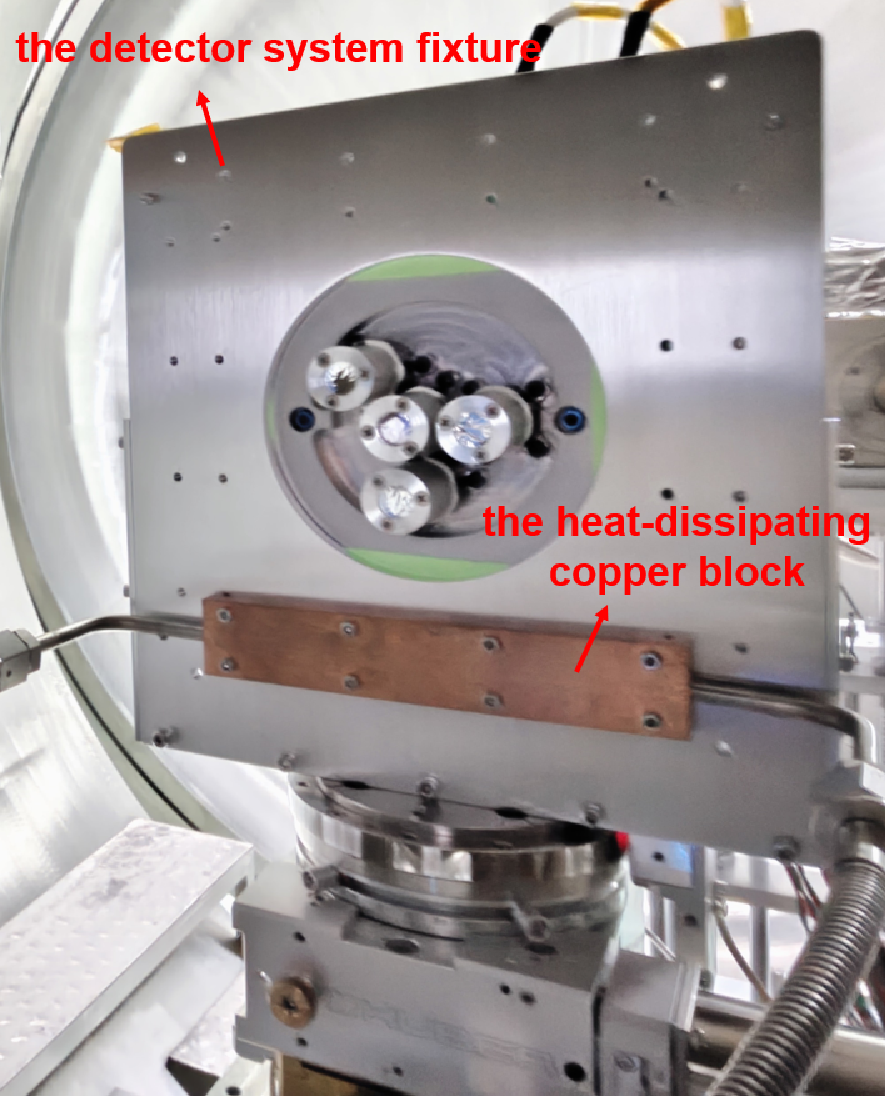}}
    \end{minipage}
\caption{Fixing the payloads on the movable stages. \textit{(left panel)} Fixed state of the mirror system: compared to the previous separate calibration for the MPO mirror, a new fixture structure \textit{(left top)} is fabricated because the optical alignment plate is installed on the MPO mirror. \textit{(right panel)} Fixed state of the detector system: the detector system fixture not only serves the functions of fixation and support but also acts as a heat plate for uniform heat dissipation.}
\label{FigFixture}
\end{figure}

\begin{figure}[ht]%
        \centering 
        \includegraphics[width=0.7\textwidth]{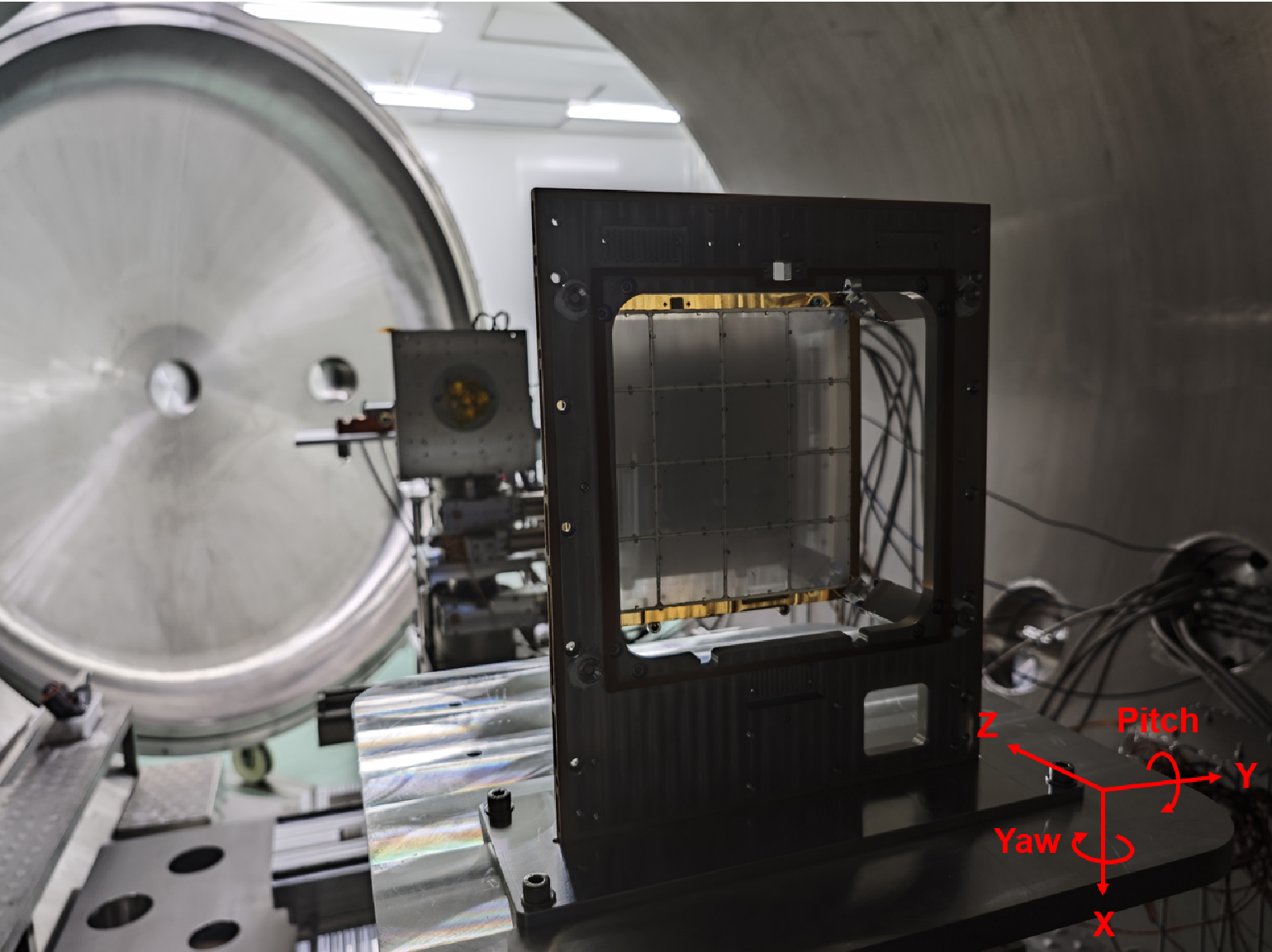}
        \caption{The MPO mirror and the detector system in the vacuum chamber to undergo joint ground calibration.}
        \label{FigVacummChamber}
\end{figure}

\begin{figure}[ht]%
        \centering 
        \includegraphics[width=1.0\textwidth]{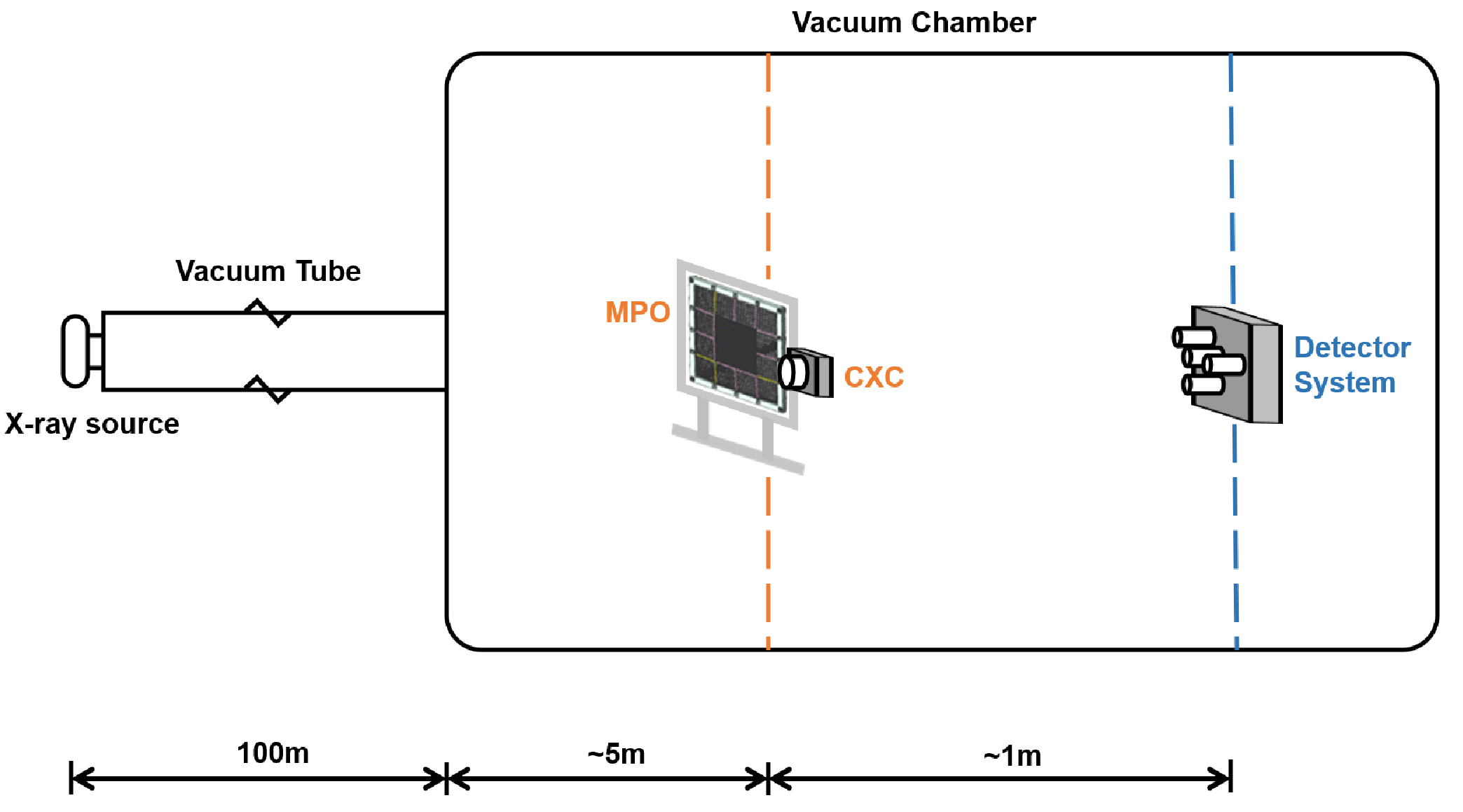}
        \caption{The relative position among the X-ray source, the MPO mirror, and the detector system. The pnCCD Color X-Ray Camera (CXC) and the MPO mirror were placed at the same distance from the X-ray source, and the CXC detector was slightly shifted toward the outside with respect to the opposite edge of the MPO mirror, which was used to monitor the beam flux passing through the MPO mirror in the subsequent calculation of the effective area.}
        \label{FigVacummPosition}
\end{figure}

Figure~\ref{FigSystem} shows the two payloads onboard CATCH-1 that required joint ground calibration. In the detector system, each SDD detector is housed within an Aluminium collimator to shield it from the photon background from the surroundings. The calibration was conducted at the 100\,m X-Ray Test Facility (100XF)~\cite{wang2023100} at the Institute of High Energy Physics (IHEP), Chinese Academy of Sciences (CAS). In 100XF, there are X-ray sources capable of generating X-ray beams of different energies and movable stages for fixing payloads, which were described in detail in Xiao et al. \cite{xiaomirror}. Figure~\ref{FigFixture} presents the fixed state of the mirror system and the detector system. For the mirror system, an optical alignment plate was installed on the MPO mirror in order to accurately align the mirror system with the detector system during the subsequent satellite integration. Therefore, a new fixture was fabricated to connect the mirror system and the movable stage. The detector system was also fixed on the movable stage by a specially fabricated fixture. As detectors need to operate at the temperature around $-$35\,$^{\circ}$C to enhance the energy resolution, a heat-dissipating copper block was mounted on the fixture to cool detectors, with pipes on both sides of the copper block to transport cold water. Hence, this fixture also serves the purpose of uniform heat dissipation. In the SDD array displayed on the right panel of Figure~\ref{FigFixture}, the central detector is the primary detector, the two detectors on the lower and right sides are cross-arms detectors, and the detector in the upper left corner is the background detector. Ultimately, the placement of the two payloads in the vacuum chamber before the formal calibration is shown in Figure~\ref{FigVacummChamber}, and the relative position among the X-ray source, the MPO mirror, and the detector system is shown in Figure~\ref{FigVacummPosition}.

\subsection{Alignment}\label{sub2}

\begin{figure}[ht]%
        \centering
        \makebox[\textwidth]{\includegraphics[width=1.0\textwidth]{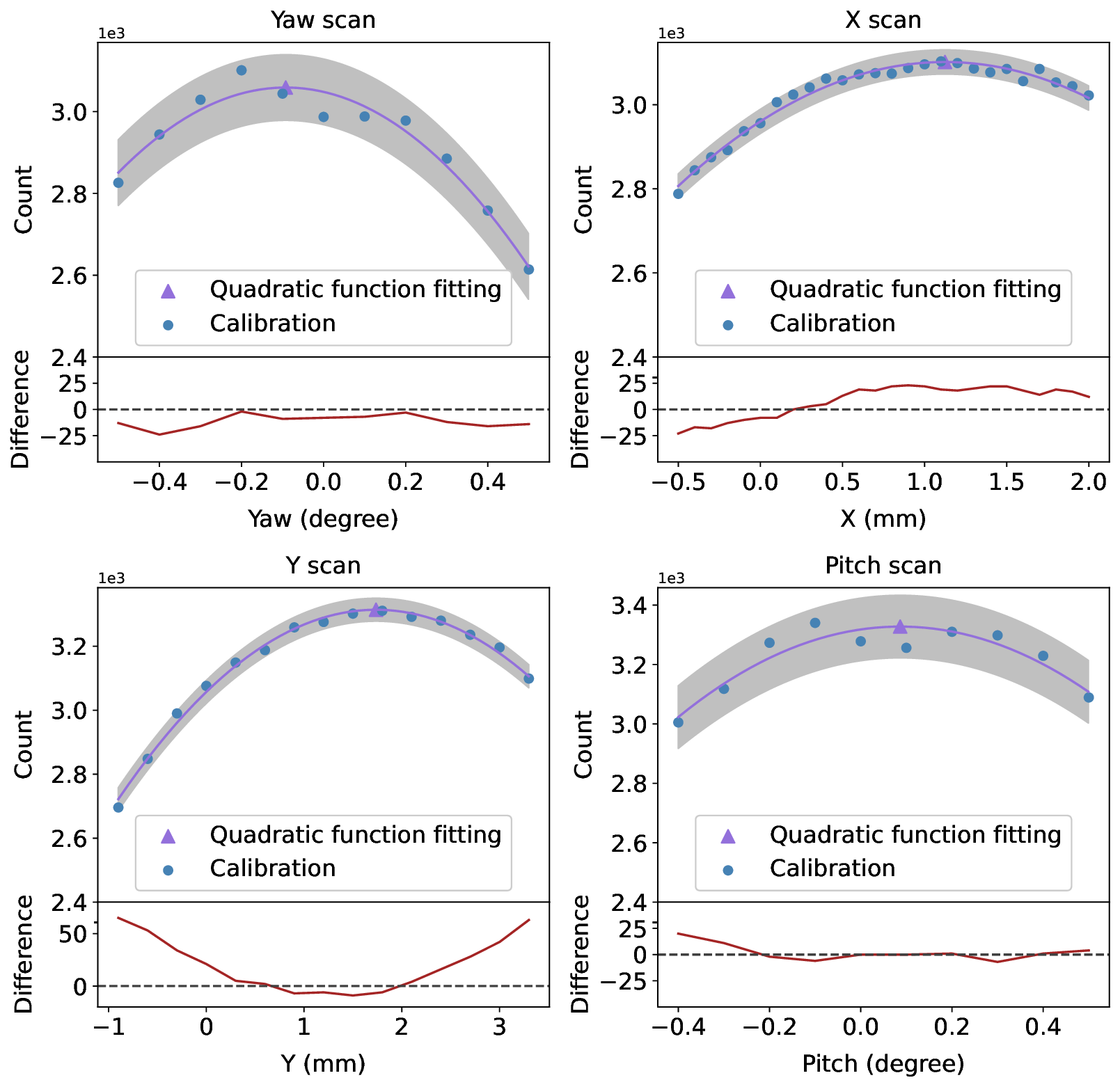}}
        \caption{Sequentially scan the four dimensions of the mirror's movable stage and determine the optimal position. Each subplot shows the counts in the primary detector in the upper part and the difference between the counts measured in the two cross-arm detectors in the lower part. The shaded area represents the 95\% confidence interval of the fitting results.}
        \label{FigScan}
\end{figure}

\begin{figure}[ht]%
        \centering 
        \includegraphics[width=0.7\textwidth]{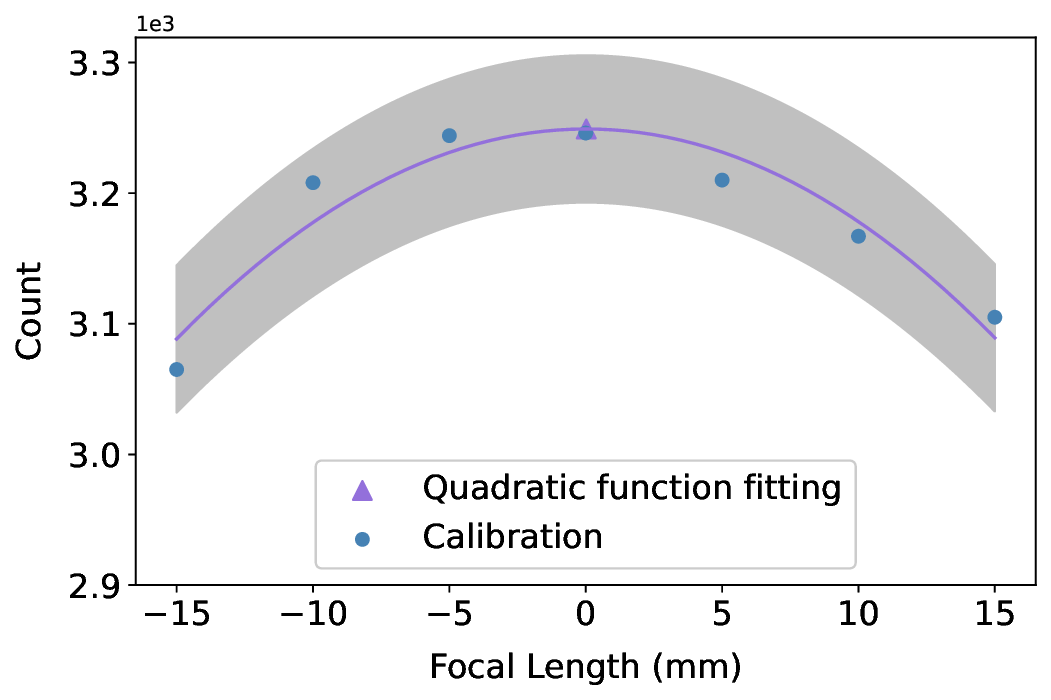}
        \caption{Searching for the focal length of the MPO mirror. The shaded area represents the 95\% confidence interval.}
        \label{FigFocus}
\end{figure}

In order to obtain accurate and effective results in the system calibration, the optimal positions of the two payloads must first be determined~\cite{rukdee2023x}. Before the vacuum chamber was closed and pumped, a coarse alignment was carried out using a laser to initially coincide the laser beam with the center of the 100\,m vacuum tube and the center of the MPO mirror, preliminarily ensuring that the focused spot falls close to the primary detector. Following that, the precise alignment was done in the ultra-high vacuum environment by scanning with X-rays at the Al-K energy near the positions of the coarse alignment to determine the optimal positions of the two payloads. The procedure for scanning the axes is to rotate or move the four axes of the mirror's movable stage in sequence: Yaw, X, Y, and Pitch, with the scanning steps of 0.1$^{\circ}$, 0.1\,mm, 0.3\,mm, 0.1$^{\circ}$, respectively. In each scan, the counts on each detector were recorded. The specific scanning range and scanning results can be found in Figure~\ref{FigScan}, which shows the counts in the primary detector and the difference between the counts measured in the two cross-arm detectors. For the counts in the primary detector, the quadratic function was used for fitting, and the optimal position for each axis was determined as where the maximum counts of the fit correspond to. In addition to presenting the calibration results and the fitting results, Figure~\ref{FigScan} also shows the 95\% confidence interval of the fitting results, indicating that the calibration results are essentially within the 95\% confidence interval. During the final scanning process of the Pitch axis, the counts in the two cross-arms detectors are basically the same due to the unique arrangement of the SDD array, indicating that the X-ray beam is very close to on-axis incidence and leading to nearly symmetric images. After determining the on-axis direction, the detector system was moved in the Z direction, i.e., the focal length direction, with a step size of 5\,mm for scanning to obtain the optimal focusing position, as shown in Figure~\ref{FigFocus}. In Figures~\ref{FigScan} and~\ref{FigFocus}, the reference zero positions correspond to the positions determined by the coarse alignment.

\subsection{Effective Area}\label{sub3}

The effective area of the system was measured at different energies. At a given energy, the effective area $A_{\text{eff}}$ (cm$^{2}$) is defined as the factor that converts the photon flux $F$ (counts~s$^{-1}$~cm$^{-2}$) of the incident X-ray beam in front of the telescope into the measured count rate $C_{\text{det}}$ (counts~s$^{-1}$) of the detector, as expressed by Equation~\ref{eqEffDefine}~\cite{cheng2023ground},
\begin{equation}
C_{\text{det}}= A_{\text{eff}} \times F.\label{eqEffDefine}
\end{equation}
To measure the photon flux, a pnCCD Color X-Ray Camera (CXC) supplied by PNDetector GmbH~\cite{CXC} with known quantum efficiency was placed on one side of the MPO mirror (shown in Figure~\ref{FigVacummPosition}) to monitor the beam flux passing through the MPO mirror as a reference. The detection energy range of the CXC detector is 0.2--10\,keV~\cite{hou}. The MPO mirror and the CXC detector are 105\,m from the source, and the light spot here is large enough to cover the MPO mirror and the CXC detector simultaneously. The effective area can be calculated using Equation~\ref{eqEff},
\begin{equation}
A_{\text{eff}}=\displaystyle{\frac{C_{\text{SDD}}}{\displaystyle{\frac{C_{\text{CXC}}}{A_{\text{CXC}} \times QE_{\text{CXC}}}}}} \times G,\label{eqEff}
\end{equation}
where $C_{\text{SDD}}$ and $C_{\text{CXC}}$ are the count rate measured by the SDD detector and the CXC, respectively. $A_{\text{CXC}}$ is the collection area of the CXC detector, and $QE_{\text{CXC}}$ is the quantum efficiency of the CXC detector. $G$ is the geometrical correction factor for the divergent X-ray beam, defined as Equation~\ref{eqG}~\cite{rukdee2023einstein},
\begin{equation}
G=\left( \frac{d_{\text{s-m}}}{d_{\text{s-m}} + d_{\text{m-SDD}}} \right) ^{2},\label{eqG}
\end{equation}
where $d_{\text{s-m}}$ is the distance between the source and the mirror, which is 105\,m, and $d_{\text{m-SDD}}$ is the distance from the MPO mirror to the SDD detector, which is taken as 993\,mm.

\begin{figure}[ht]%
        \centering
        \includegraphics[width=0.75\textwidth]{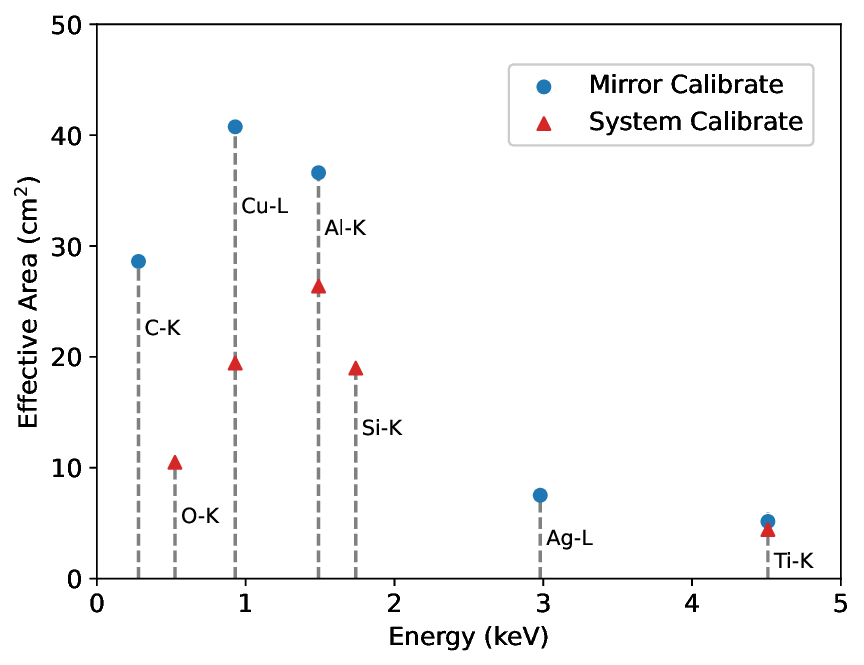}
        \caption{Comparison of the effective area of the combination of the MPO mirror and the SDD detector obtained from the system calibration and the effective area obtained from the previous separate calibration for the MPO mirror.}
        \label{FigSystemEffArea}
\end{figure}

\begin{figure}[ht]%
        \centering 
        \includegraphics[width=0.6\textwidth]{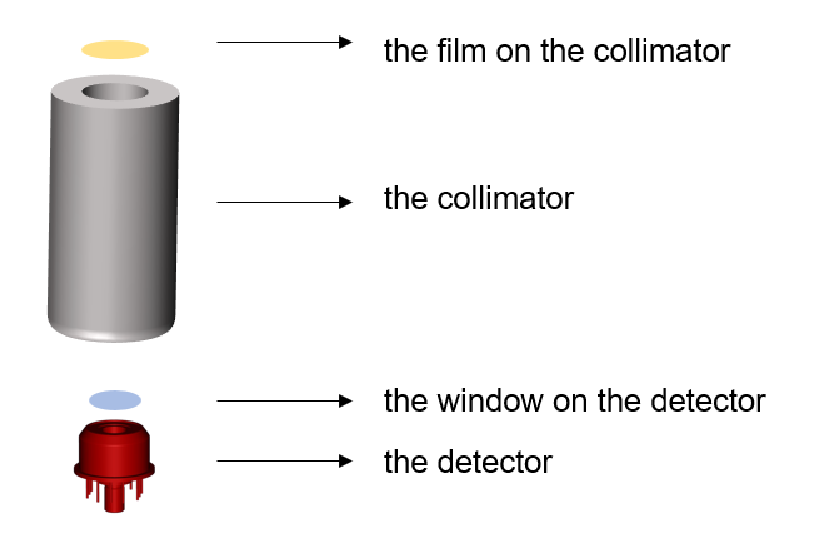}
        \caption{The exploded view of one detector and its collimator. Before the focused photon reaches the detector, it will pass through the film on the collimator, the collimator, and the detector window.}
        \label{FigDifference}
\end{figure}

The effective area was measured at five energies: O-K (0.525\,keV), Cu-L (0.93\,keV), Al-K (1.49\,keV), Si-K (1.74\,keV), and Ti-K (4.51\,keV). The effective area at each energy is listed in Table~\ref{effective_area}. Figure~\ref{FigSystemEffArea} compares the results of this calibration with the previous separate calibration for the MPO mirror. The difference between the two calibrations is that this calibration provides the effective area when the MPO mirror is used in combination with the SDD detector, while the previous calibration provided the effective area of the MPO mirror alone. In the previous calibration of the MPO mirror, the CXC detector with known quantum efficiency was placed at the position of the SDD detector in this calibration, i.e. the focal plane. The effective area was calculated using the flat field signal when the MPO mirror was moved away from the front and the focusing signal when the MPO mirror was aligned~\cite{xiaomirror}. The difference in results between the two calibrations can be attributed to three factors (see Figure~\ref{FigDifference}): the transmission of the film on the collimator, the transmission of the detector window, and the quantum efficiency of the detector itself. It can be seen that the choice of energies for the two calibrations is different, which is due to the different available targets in the calibration hall during the two calibrations. The selection of energy is based on being within or near the target energy range for the scientific detection by CATCH-1, i.e. 0.5--4\,keV.

\begin{table}[ht]
\renewcommand\arraystretch{1.5}
\caption{Summary of the effective area measured at each energy line. }\label{effective_area}%
\begin{tabular}{p{2cm}<{\centering}p{3cm}<{\centering}p{3cm}<{\centering}}
\toprule
Name  &   Energy (keV)     &  Effective Area (cm$^{2}$) \\
\midrule
O-K  & 0.525  &  10.50   \\
Cu-L  & 0.93  &  19.44   \\
Al-K  & 1.49  &  26.38   \\
Si-K  & 1.74  &  18.99   \\
Ti-K  & 4.51  &  4.42   \\
\botrule
\end{tabular}
\end{table}

\section{Calibration Database}\label{sec3}

\begin{figure}[ht]%
        \centering
        \includegraphics[width=1.0\textwidth]{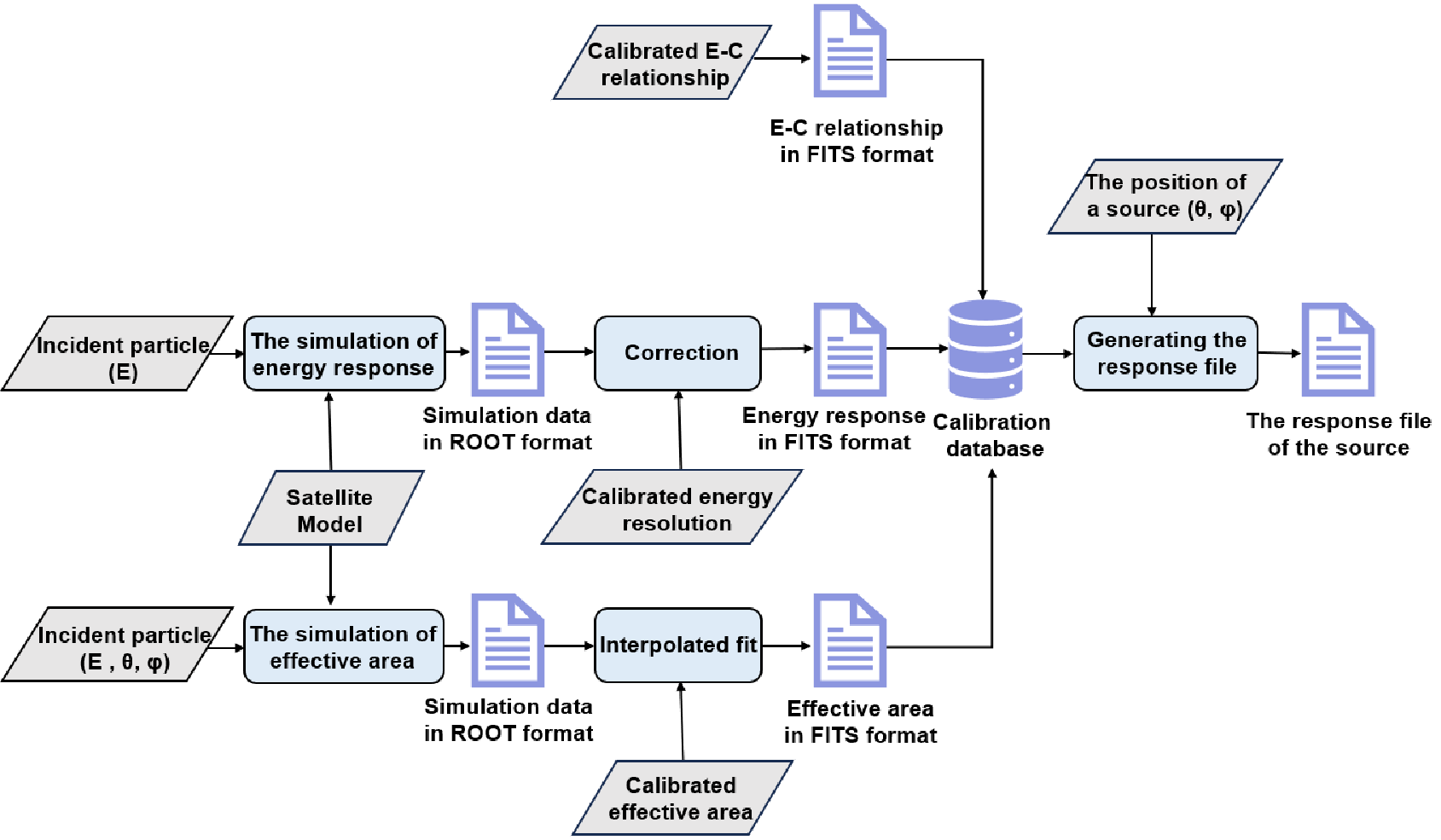}
        \caption{The flowchart to establish the ground calibration database of CATCH-1.}
        \label{FigArchitecture}
\end{figure}

The establishment of the ground calibration database is a fundamental work prior to the launch of an X-ray astronomy satellite. The calibration database of CATCH-1 consists of two major data products: Redistribution Matrix File (RMF) and Ancillary Response File (ARF). These two data products involve three key relationships: energy-channel (E-C) relationship, energy response matrix, and effective area. The specific process for establishing the ground calibration database is illustrated in Figure~\ref{FigArchitecture}. 

The E-C relationship is directly obtained through the calibration of the detector system, with detailed results in Li et al.~\cite{lidetector}. To enhance reliability and redundancy, CATCH-1 is equipped with two sets of electronic readout systems, each designed and calibrated separately. There are 16,384 energy channels in the main system and 2,048 channels in the backup system. Due to the variation of the E-C relationship with environmental temperature, high voltage, etc., the Pulse Invariant (PI) is commonly defined for convenience. CATCH-1 has a total of 970 PI, covering the energy range of 0.3--10\,keV, with a channel width of 0.01\,keV.

\begin{figure}[ht]%
        \centering
        \includegraphics[width=0.7\textwidth]{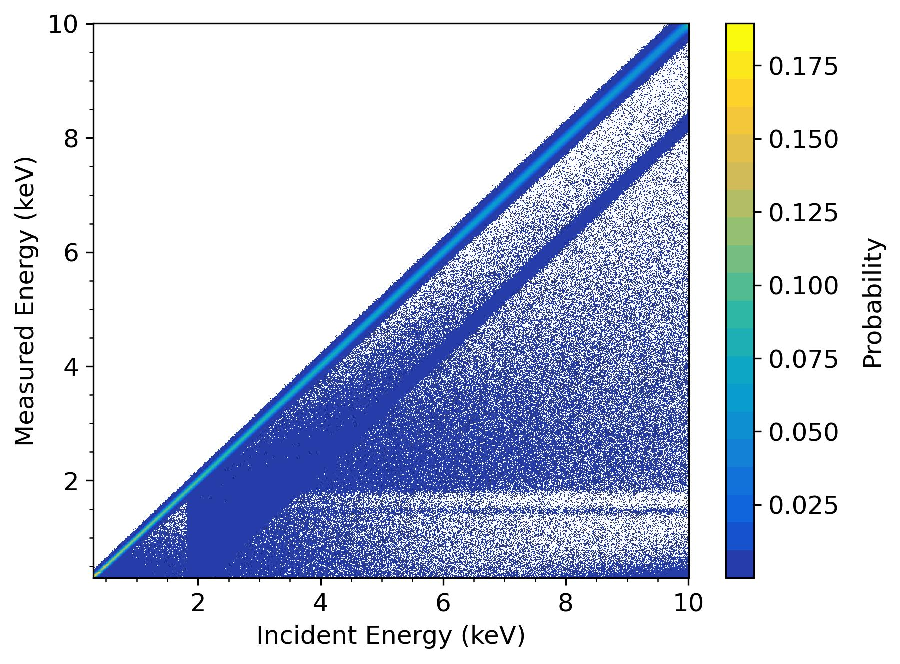}
        \caption{The Energy response matrix of the primary detector in the primary electronic readout systems.}
        \label{FigEnergyResponse}
\end{figure}

The energy response matrix and effective area were obtained through computer modeling and Monte Carlo simulations. The simulation work was performed using the Geant4 Monte Carlo toolkit (version 4.10.06.p03), which allows for the construction of a complete satellite model and simulation of particle transport processes in matter~\cite{1geant4, 2geant4, 3geant4, huangsimulation}. The energy deposition distributions of incident photons at various energies in each detector of CATCH-1, i.e. energy responses, can be obtained from simulation. These energy responses, after convolving with the calibrated energy resolution of each detector and normalization, yield the data for the energy response matrix of CATCH~\cite{zhang2019energy, guo2020energy}. Figure~\ref{FigEnergyResponse} shows the energy response matrix of the primary detector in the primary electronic readout systems. The target energy range for scientific detection by CATCH is 0.5--4\,keV; however, the simulated range of incident energy and the recorded range of the deposition energy are both 0.3--10\,keV, considering the detection capability of the SDD detector and the focusing capability of the MPO mirror. It can be observed that when the energy of the incident photon is below the K-shell absorption edge of Si, i.e. 1.84\,keV, only the full energy peak is present. When the energy is higher than 1.84\,keV, a characteristic X-ray peak of Si emerges, accompanied by the appearance of an escape peak due to the escape phenomenon. The relative positions of the escape peak and full energy peak remain constant, and the absolute position of the characteristic X-ray peak remains unchanged. Through simulations, it was found that the energy response matrix is not affected by the off-axis angle and off-axis direction of incident particles, hence the position variation was not considered during the establishment of the energy response matrix.

\begin{figure}[htbp]
  \centering
  \begin{minipage}[t]{0.9\linewidth} 
      \centering
      \includegraphics[width=9cm]{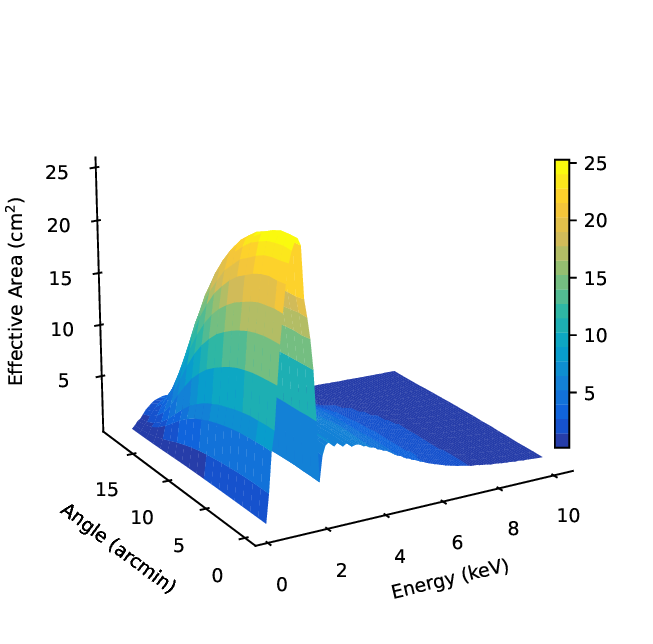}
  \end{minipage}
  \begin{minipage}[t]{0.9\linewidth}
      \centering
      \includegraphics[width=12cm]{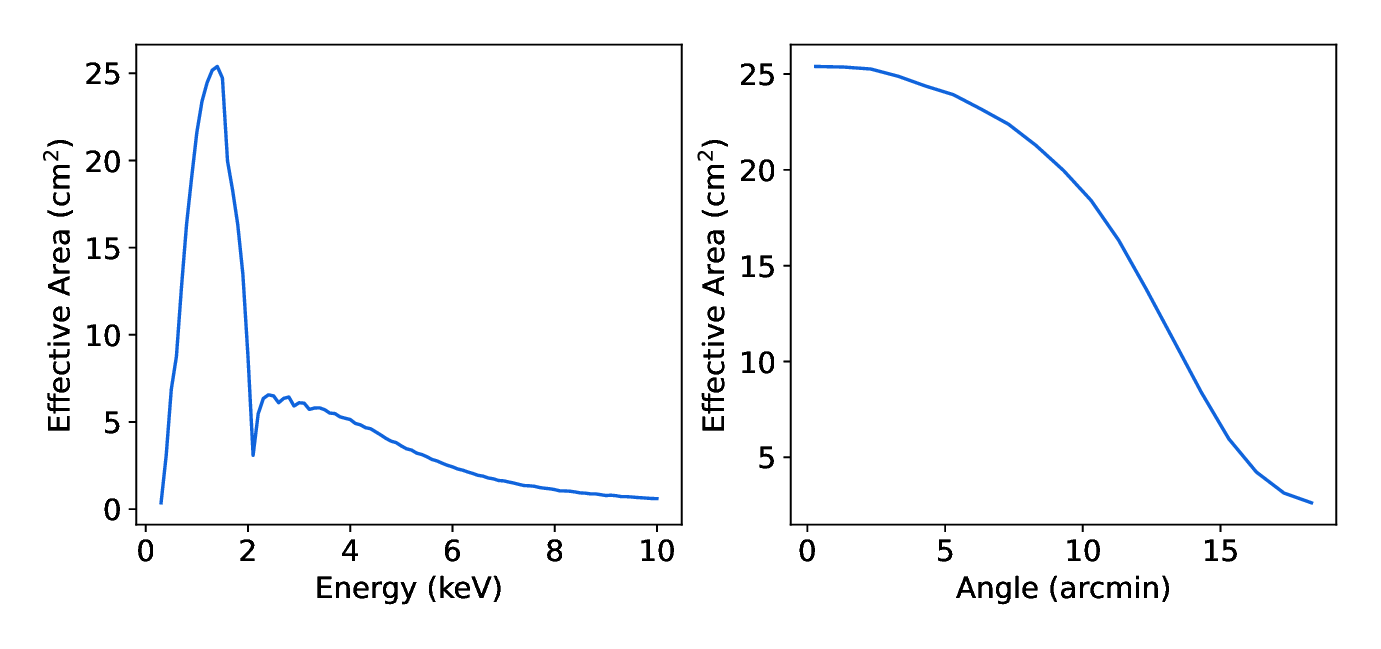}
  \end{minipage}
  \caption{\textit{(top panel) The variation of the effective area with the energy and off-axis angle of the incident photons when the off-axis direction is \ang{0}. \textit{(bottom left)} The variation of the effective area with the energy when the off-axis angle is \ang{0}. \textit{(bottom right)} The variation of the effective area with the off-axis angle when the energy is 1.4\,keV.}}
  \label{Figene_angle_area}
\end{figure}

The effective area is influenced by the energy, off-axis angle, and off-axis direction of incident photons. Through simulation, the effective area of incident photons at various energies and positions can be obtained. The simulated effective area is consistent with the calibrated values presented in Section~\ref{sub3}. Therefore, by integrating the simulated effective area data with the calibrated effective area data, the effective area at any energy and any position can be generated through three-dimensional interpolation fitting. The top panel in Figure~\ref{Figene_angle_area} presents the variation of the effective area of the primary detector with the energy and off-axis angle of incident photons when the off-axis direction is \ang{0}. To provide a clearer depiction of the trend in effective area variation, the bottom left panel displays how the effective area changes with energy when the off-axis angle is \ang{0}. It can be seen that the effective area reaches its maximum at 1.4\,keV, decreases rapidly thereafter, starts to increase again at 2.1\,keV, and then gradually decreases. The variation in the effective area is related to the reflectivity of the coating on the reflective surface of the focusing mirror, which varies with the material and thickness of the coating. There is a layer of 20\,nm Iridium coating on the sides of the micro pores of the MPO mirror on CATCH-1. Additionally, as mentioned in Section~\ref{sub3}, the system-level effective area is also affected by the transmission of the film on the collimator, the transmission of the detector window, and the quantum efficiency of the detector itself. The bottom right panel displays the effective area's dependence on the off-axis angle for an energy of 1.4\,keV. With the increase of the off-axis angle, the effective area decreases gradually. 

The established ground calibration database can achieve the functionality of generating response files for any incident source at a given position. The ground calibration database can be utilized to assess CATCH-1's in-orbit observational capabilities and provide support for the workflow of data analysis algorithms and software. Following further refinement, it will play a crucial role in future scientific analysis during in-orbit operations.

\section{Ground network}\label{sec4}

\begin{figure}[ht]%
        \centerline{\includegraphics[width=1.0\linewidth]{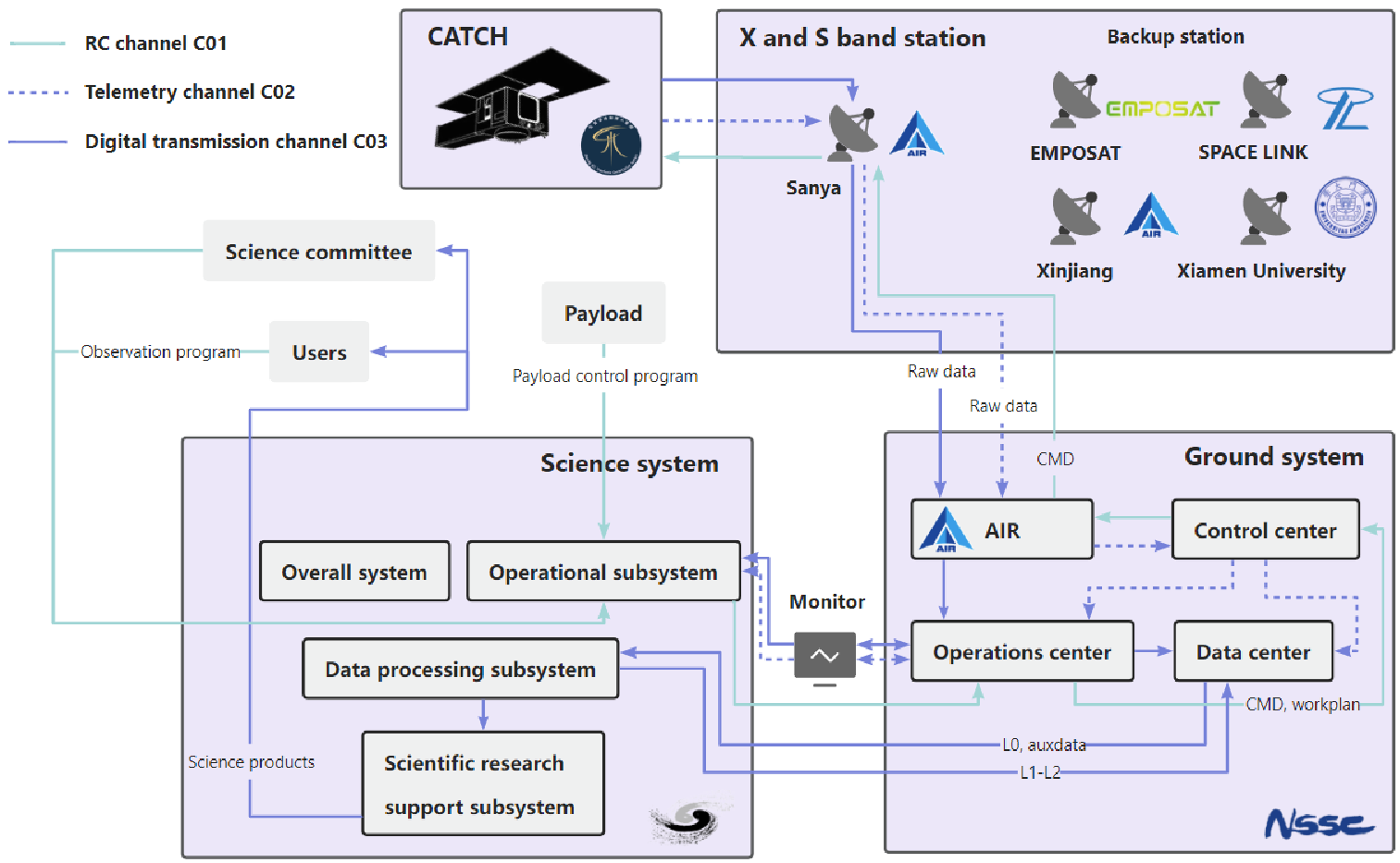}}
        \caption{CATCH ground network.}
        \label{FigGroundNetwork}
\end{figure}

Communication between the ground and the in-orbit CATCH-1 is via the primary ground station in Sanya (China) and other backup ground stations, such as EMPOSAT, SPACE LINK, Xinjiang, and Xiamen University. There are three channels for transmitting instructions or data. The uplink instructions are transmitted through the RC channel (C01), while the engineering data and scientific data generated by the in-orbit CATCH-1 are sent back to the ground through the telemetry channel (C02) and the digital transmission channel (C03). The transmission of the three channels among CATCH-1, the ground station, the ground system, and the science system is shown in Figure~\ref{FigGroundNetwork}~\cite{2021SPIE11821E..0MR}.

The CATCH-1 science committee and users can submit three types of observation proposals to the science system managed by IHEP based on their scientific objectives, including long-term (once every 3 months), regular (daily), and Target of Opportunity (ToO). Additionally, payload providers can submit payload control programs. Subsequently, the ground system managed by the National Space Science Center (NSSC) will generate the corresponding commands, which will then be sent to CATCH-1 via the Sanya ground station.

Following the scheduling by NSSC, the raw data obtained by the in-orbit CATCH-1 will be transmitted to NSSC via the Sanya ground station. The NSSC will decode the data to provide real-time monitoring and generate 0D-level engineering data, 0B-level scientific data, and auxiliary data, which will subsequently be distributed to IHEP. IHEP will further process the data to generate science data products for user utilization.

\section{Satellite integration}\label{sec5}

The results in Figure~\ref{FigFocus} indicate that the counts within the geometry range of the primary detector are not sensitive to changes in the focal length within $\pm$10\,mm. In the previous separate calibration for the MPO mirror using the CXC detector, the imaging capability of the CXC detector allowed us to more accurately determine the optimal position in the focal length direction through the angular resolution of the focal spot imaging. Additionally, the distance from the upper surface of the detector system to the sensitive layer of the SDD detector is difficult to measure. Therefore, we used the distance of 993\,mm between the MPO mirror to the CXC detector measured in the previous calibration after the alignment was performed as the distance from the MPO mirror to the SDD detector for this calibration. By calculating the correction of 9.43\,mm required for the divergent X-ray beam based on this distance, we extended the distance between the MPO mirror and the SDD detector by 9.43\,mm to achieve the optimal positions under parallel light conditions.

\begin{figure}[ht]
\centering  
\subfigure{
\includegraphics[width=6.15cm,height = 5.15cm]{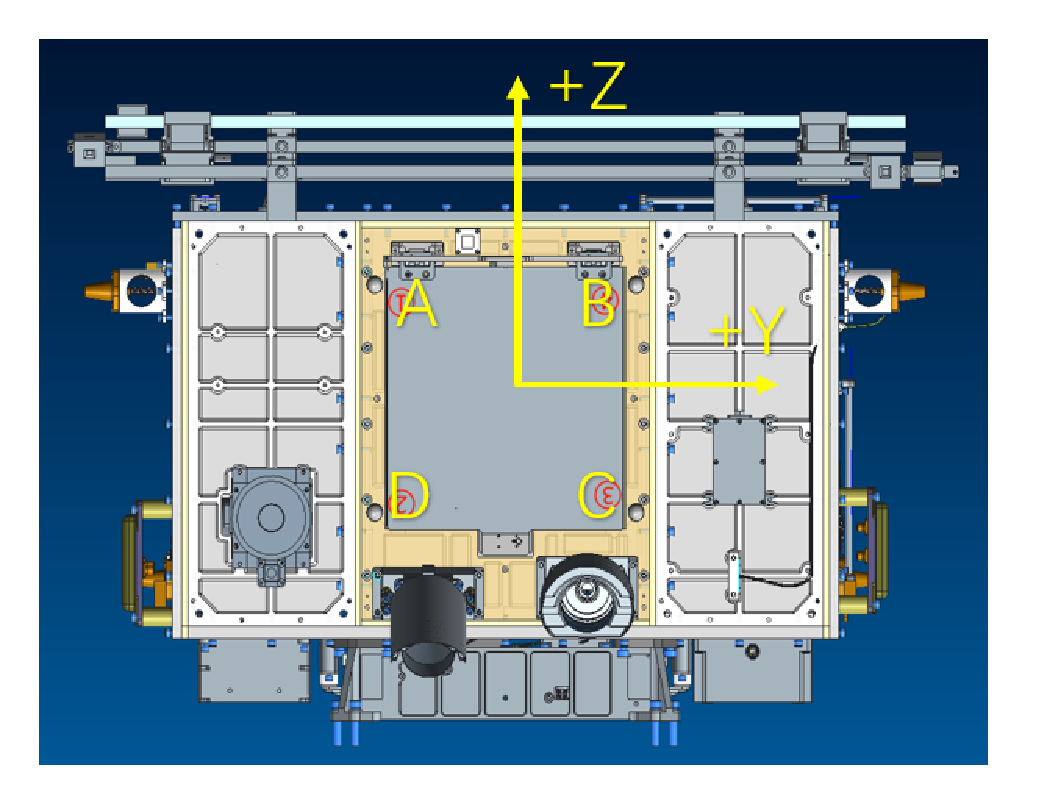}}
\subfigure{
\includegraphics[width=6.15cm,height = 5.3cm]{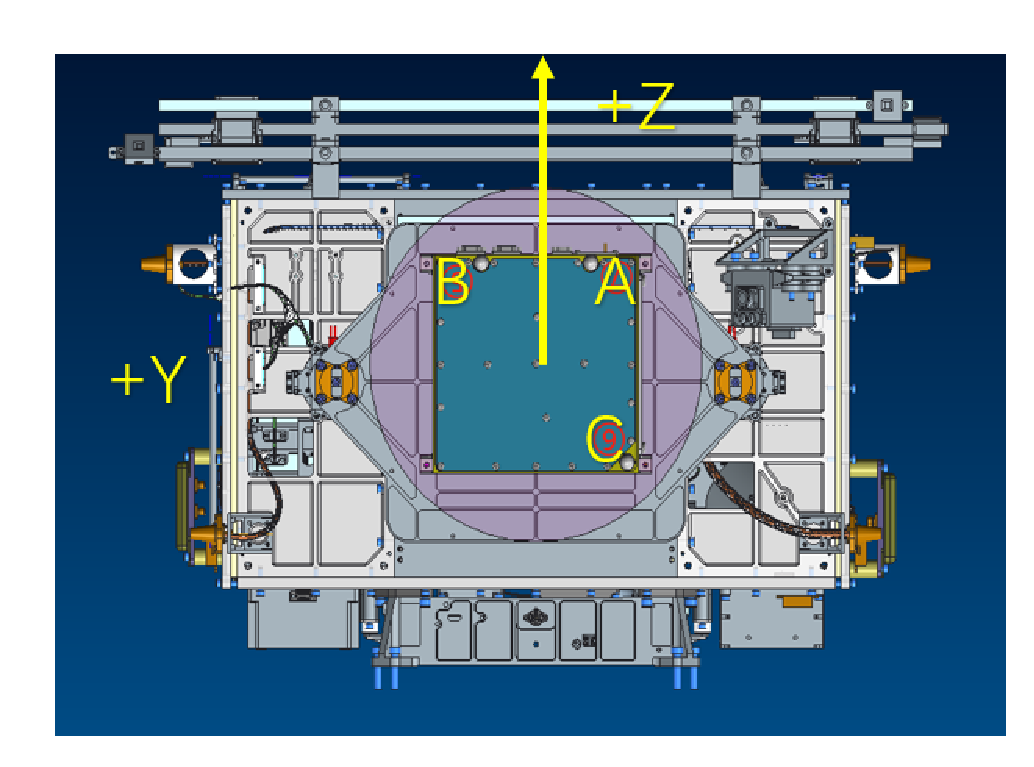}}
\subfigure{
\includegraphics[width=11.15cm,height = 6.0cm]{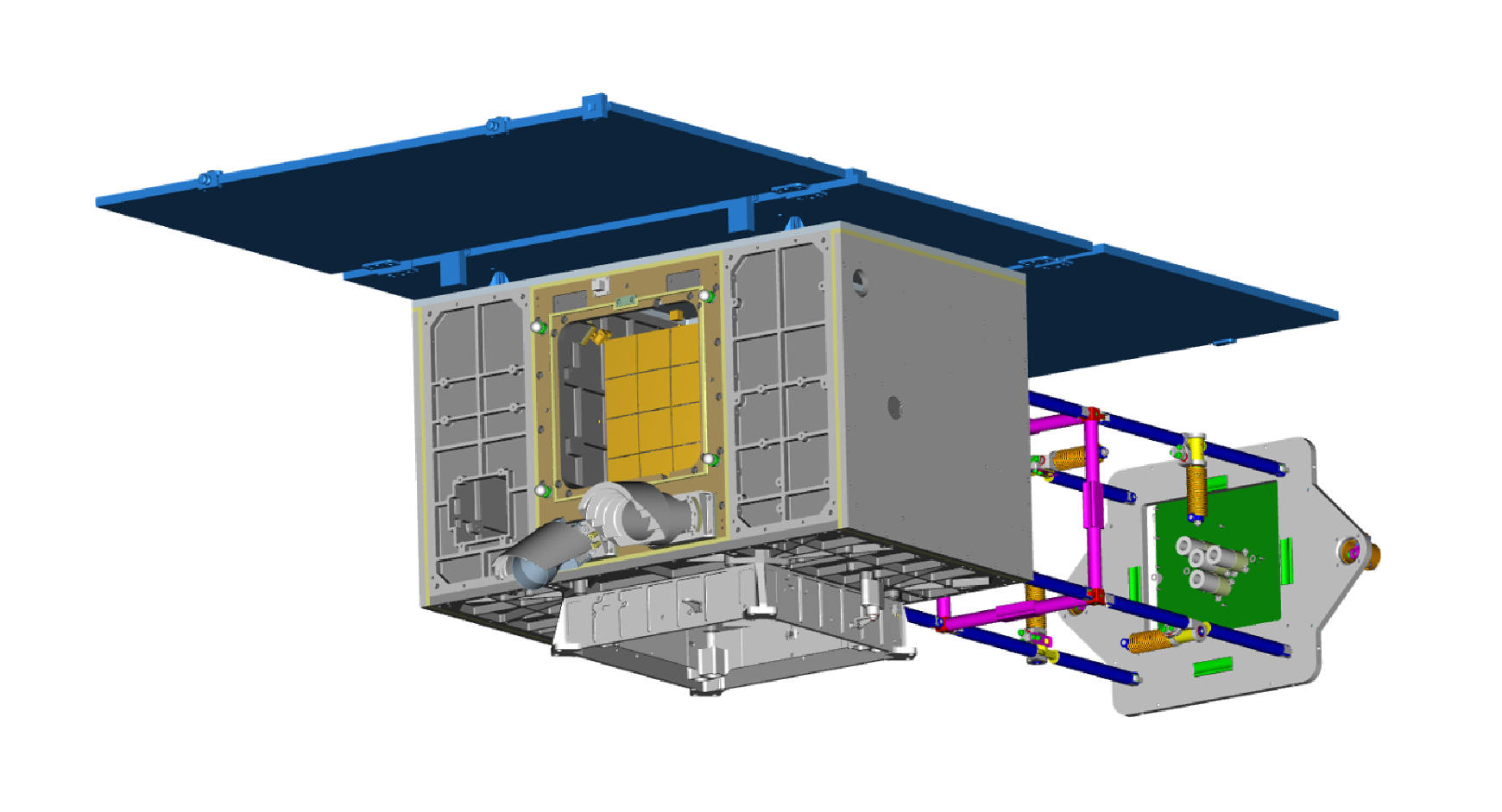}}
\caption{\textit{(top panel)} The positions of the target balls in the satellite coordinate system: \textit{(top left)} on the mirror system, there are four target balls labeled A, B, C, and D; \textit{(top right)} on the detector system, there are three target balls labeled A, B, and C. \textit{(bottom panel)} The placement of the mirror system and the detector system in the satellite.}
\label{FigPosition}
\end{figure}

\begin{figure}[ht]%
        \centering
        \includegraphics[width=0.7\textwidth]{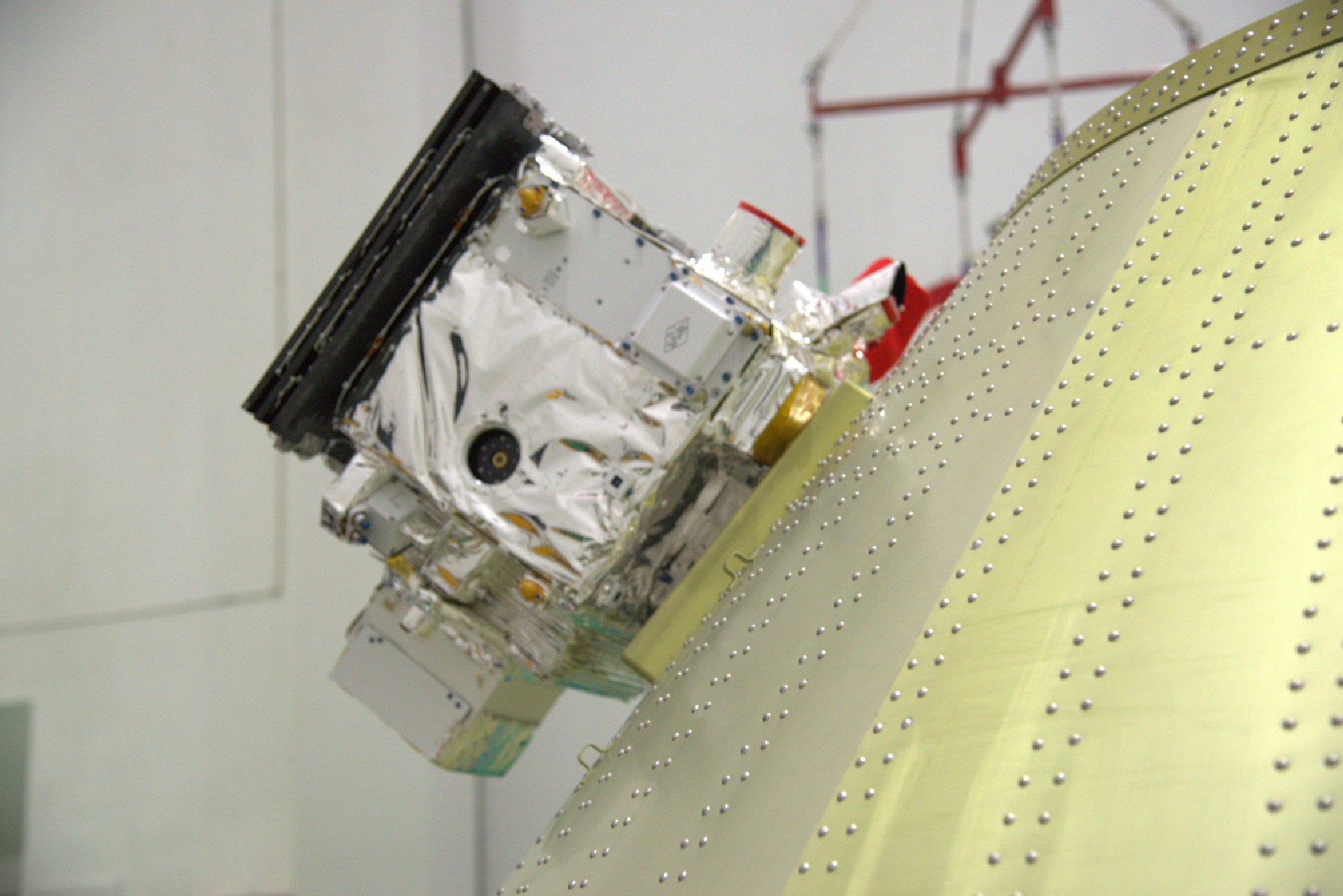}
        \caption{The fully integrated CATCH-1 has been mounted on the rocket and is prepared for launch.}
        \label{FigCATCH_1}
\end{figure}

Subsequently, the vacuum chamber was opened and the team from the Innovation Academy for Microsatellites of CAS (IAMCAS) entered it and measured the coordinates of the target balls on the mirror system and the detector system (see Figure~\ref{FigPosition}). By combining the optical reference information provided by NAOC, they measured the coordinates of the optical alignment plate and target balls on the mirror system at IAMCAS. These two measurements were used to determine the relative position between the detector system and the optical alignment plate, which is required for the satellite integration. Ultimately, they accomplished the full integration of CHATC-1. The final configuration of the satellite is shown in the bottom panel of Figure~\ref{FigPosition}, with the mirror system located in the front of the platform and the detector system located at the end of the deployable mast. The final state of the whole satellite is shown in Figure~\ref{FigCATCH_1}, with the solar panels and the deployable mast folded to minimize payload volume during launch.

\section{Summary}\label{sec6}

This paper provides a comprehensive technical overview of the pre-launch preparations for the CATCH pathfinders, using CATCH-1 as a case study. The discussed works include system calibration, calibration database, ground network, and satellite integration. In order to assess the scientific detection capability for the combination of the MPO mirror and the SDD detectors on CATCH-1, the system calibration was performed, and specialized fixtures were fabricated for this calibration. The effective area at five energies was obtained in calibration, with the largest effective area of 26.38\,cm$^{2}$ at 1.49\,keV. The results from the system calibration combined with those from Monte Carlo simulations allow the establishment of a ground calibration database. This paper presents the process of establishing the ground calibration database and the energy response matrix and effective area in the ground calibration database. The system calibration also played a crucial role in the full integration of the satellite. During calibration, the MPO mirror and the SDD detectors were adjusted to be in the optimal position. The team from IAMCAS used the measured position information of each target ball to ultimately achieve the accurate integration of the satellite. Additionally, the collaborative process of the ground network during satellite operations in orbit is introduced. 

\section*{Acknowledgments}

We would like to express our gratitude to all colleagues in the CATCH team for their contributions. This work is supported by the National Natural Science Foundation of China (NSFC) under the Grant Nos. 12122306, 12003037, and 12173056, and the Strategic Priority Research Program of the Chinese Academy of Sciences under Grant No. XDB0550300. Their support made this work possible.

\section*{Author Contributions}

Yiming Huang and Jingyu Xiao co-authored the paper and made equal contributions. Qianqing Yin led the system calibration experiment, Yusha Wang and Chen Zhang participated in the experimental design, Zijian Zhao performed the experimental operation, and Yiming Huang and Jingyu Xiao analyzed the calibration data. Qingchang Zhao led the establishment of the ground calibration database, and Yiming Huang provided the simulation data. Xiang Ma, Yue Huang, and Shujie Zhao are jointly responsible for the coordination and communication of the ground network. Heng Zhou and Yong Yang led the integration of the satellite. Lian Tao and Shuang-Nan Zhang proposed the CATCH mission and assisted with the critical revision of the article. All authors reviewed the manuscript and contributed to the pre-launch work of CATCH-1.

\section*{Funding}

This work is supported by the National Natural Science Foundation of China (NSFC) under Grant Nos. 12122306, 12173056 and 12003037, as well as support from the Strategic Priority Research Program of the Chinese Academy of Sciences under Grant No. XDB0550300.

\section*{Conflict of interest}

The authors declare that they have no conflict of interest.


\bibliography{sn-bibliography}

\end{document}